\documentclass[twocolumn,showpacs,preprintnumbers,amsmath,amssymb,epsf]{revtex4-1}

\usepackage{graphicx,epsfig}
\usepackage{dcolumn}
\usepackage{bm}

\begin{document}


\title{Simulations of Collisionless Perpendicular Shocks in Partially Ionized Plasmas}
\author{Yutaka Ohira}
\affiliation{Department of Physics and Mathematics, Aoyama Gakuin University, 5-10-1 Fuchinobe, Sagamihara 252-5258, Japan}


\begin{abstract}
Perpendicular collisionless shocks propagating into partially ionized plasmas are investigated 
by two-dimensional hybrid particle simulations. 
It is shown that some neutral particles leak into the upstream region from the downstream region, 
the leaking neutral particles become pickup ions in the upstream region and modify the shock structure, 
the pickup ions are preferentially accelerated, and plasma instabilities are excited by the pickup ions in the upstream and downstream regions.
\end{abstract}

\maketitle

Collisionless shocks have long been regarded as efficient 
cosmic ray accelerators in the universe \cite{axford77}. 
In fact, observations of supernova remnants (SNRs) provide the evidence that 
electrons and ions are accelerated to highly relativistic energy \citep{koyama95}. 
However, shock structures and injection to the shock acceleration 
have not been understood completely. 
Previous studies by particle simulations have addressed only shocks in 
fully ionized plasmas. 

The interstellar medium is not always fully ionized plasmas.
The existence of neutral particles around collisionless shocks 
has been identified in many SNRs 
from observations of H$\alpha$ emission \citep{chevalier78}. 
Recently, some authors proposed important effects of the neutral particles 
on collisionless shocks and particles accelerations 
\citep{ohira09,ohira10,ohira12,blasi12,morlino13}.
One of the most interesting results is that some neutral particles 
leak into the shock upstream region from the downstream region.  
The leaking neutral particles change the shock structure and 
the energy spectrum produced by the shock acceleration \citep{ohira12,blasi12}. 
However, it has not been demonstrated in ab initio particle simulations so far.

In this Letter, we present the first hybrid simulations of nonrelativistic collisionless 
perpendicular shocks propagating into partially ionized plasmas. 
In the hybrid code, ions are treated as nonrelativistic particles 
and electrons are a mass-less fluid to satisfy the charge quasi neutrality, 
that is, the motion of electrons is not solved. 
The hybrid code computes the motion of ions as coupled to 
Maxwell's equations in the low-frequency limit \citep{lipatov02}. 
In addition, we solve charge exchange of hydrogen atoms with protons and collisional ionization of hydrogen atoms 
with electrons, protons and hydrogen atoms, and the motion of hydrogen atoms as the free streaming in this Letter. 
We take into account the velocity dependence of their cross sections \citep{heng07}. 
At each time step, we calculate above processes of each hydrogen atom as follows. 
First of all, we calculate relative velocities between each hydrogen atom 
and all particles existing in the same cell, 
$v_{{\rm rel},s,ij}=|\vec{v}_{{\rm H},i}-\vec{v}_{s,j}|$, 
where the subscript $s$ represents particle species (hydrogen, proton and electron).  
$\vec{v}_{{\rm H},i}$ and $\vec{v}_{{\rm s},j}$ are the velocity of $i$-th hydrogen atom and 
the velocity of $j$-th particle of $s$. 
Here, we assume that the velocity of all electrons is the same as the mean velocity of 
that of protons existing in the same cell 
and the number of electrons is the same as that of protons. 
Then, we calculate all reaction rates of each hydrogen atom with all particles existing in the same cell.
Probabilities of all the reactions are obtained by multiplying all the reaction rates by the time step.
Finally, by using a random number, we decide with which particle and by which reaction 
each hydrogen atom becomes a proton or it still remains the hydrogen atom. 
For change exchange, the interacting proton becomes a hydrogen atom. 

The ratio of the charge exchange frequency, $\nu=n\sigma v_{\rm rel}$, 
to the cyclotron frequency, $\Omega_{\rm cp}$, is given by 
\begin{eqnarray}
\frac{\nu}{\Omega_{\rm cp}}\approx 10^{-5} \left( \frac{\sigma v_{\rm rel}}{10^{-7}~{\rm cm^3/s}} \right)\left( \frac{n}{1~{\rm cm^{-3}}} \right) \left( \frac{B}{3~{\rm \mu G}} \right)^{-1}
\label{pgamma}
\end{eqnarray}
where $n, \sigma, v_{\rm rel}$ and $B$ are the number density, 
the cross section of charge exchange, 
the relative velocity, and the magnetic field strength, respectively. 
The reaction rate coefficient, $\sigma v_{\rm rel}$, is normalized by 
the typical value for $v_{\rm rel}=2000~{\rm km/s}$, 
and the number density and the magnetic field strength are normalized by 
typical values of the interstellar medium. 
In order to reduce the computational cost, we set 
$\nu/\Omega_{\rm cp}\approx10^{-2}$, 
that is, all the cross sections or $n/B$ are enhanced by 
a factor of $10^{3}$, but all the reaction rates are 
still much smaller than $\Omega_{\rm cp}$.

We set a two-dimensional simulation box in the $xy$ plane 
with the periodic boundary condition in the $y$ direction. 
Simulation particles are injected at the left boundary, $x=0$, 
and reflect at the right boundary, 
$x=20000~c / \omega_{\rm pp}$, where $c$ and $\omega_{\rm pp}$ 
are the speed of light and plasma frequency of protons, respectively.
The simulation box size is $L_x \times L_y = 20000~c / \omega_{\rm pp} \times 400~c / \omega_{\rm pp}$.
The cell size and time step are 
$\Delta x = \Delta y = 0.5~c / \omega_{\rm p}$ and $\Delta t=0.0125~\Omega_{\rm cp}^{-1}$, respectively.
Initially, the number of simulation particles are 16 in each cell 
for protons and hydrogen atoms and 
the magnetic field is taken to be spatially homogenous, pointing 
in the $y$ direction, $\vec{B}=B_0 \vec{e_y}$. 
We have also performed a simulation for the case of the uniform 
magnetic field of the $z$ direction, $\vec{B}=B_0 \vec{e_z}$. 
Because the results are essentially the same as that of $\vec{B}=B_0 \vec{e_y}$, 
we show only the case of $\vec{B}=B_0 \vec{e_y}$. 
The plasma parameters are as follows: 
The upstream ionization fraction is $0.5$, the drift velocity 
of the $x$ direction is $v_{\rm d}=10~v_{\rm A}$, 
where $v_{\rm A}=B_0/\sqrt{4\pi \rho_{\rm p,0}}$ is the Alfv{\'e}n velocity 
and $\rho_{\rm p,0}$ is the proton mass density in the upstream region, 
the ratio of the particle pressure to the magnetic pressure is 
$\beta_{\rm p}=\beta_{\rm H}=0.5$ for protons and hydrogen atoms.
We have to specify the velocity scale to calculate charge exchange 
and collisional ionization, 
so that we set $v_{\rm d}=10~v_{\rm A} = 2000~{\rm km/s}$ 
to reproduce the typical shock velocity of young SNRs. 

The hybrid code can not solve the behavior of electrons exactly. 
Moreover, the electron heating in the collisionless shock has not 
been understood yet \citep{ohira07}.
For simplicity, we assume $T_{\rm e}=0$ in this Letter, 
where $T_{\rm e}$ is the electron temperature. 
This assumption does not significantly change our results because 
the ionization by electrons is subdominant or comparable to the ionization by protons for $v_{\rm rel}\gtrsim 2000{\rm km/s}$ \citep{heng07}. 
Note that electrons can ionize hydrogen atoms even for $T_{\rm e}=0$ 
because of a nonzero relative velocity. 
The dependence on $T_{\rm e}$ will be addressed in future work. 

\begin{figure}
{\centering
{\includegraphics[width=8.6cm]{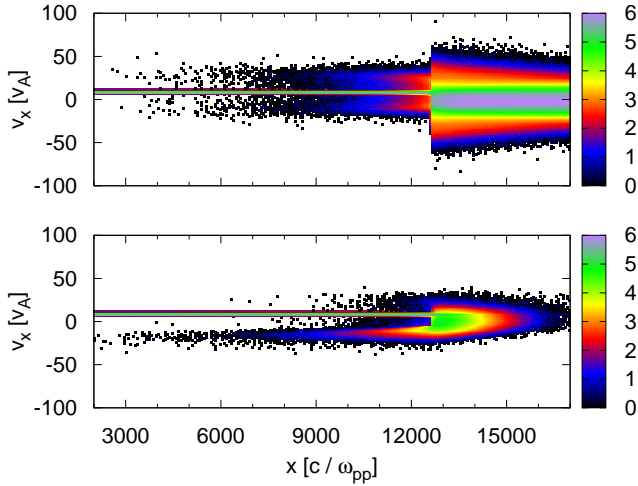}}
\par}
\caption{Phase space plots of protons (top) and hydrogen atoms (bottom) at $t=2000~\Omega_{\rm cp}^{-1}$. The color shows the phase space density in logarithmic scale.}
\label{f1}
\end{figure}
\begin{figure}
{\centering
{\includegraphics[width=8.6cm]{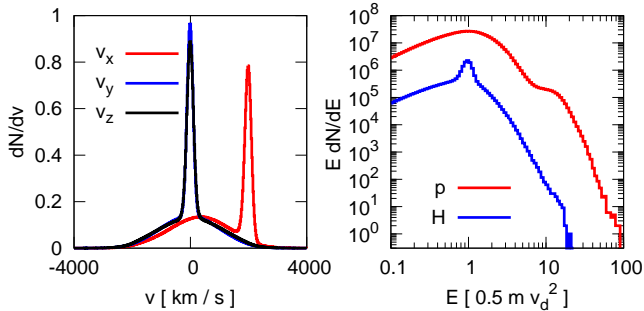}}
\par}
\caption{Velocity distribution of hydrogen atoms (left) and energy spectra (right) in the downstream region, 
$12700~c / \omega_{\rm pp} \leq x \leq17000~c / \omega_{\rm pp}$, 
at time $t=2000~{\Omega_{\rm cp}}^{-1}$.}
\label{f2}
\end{figure}
\begin{figure}
{\centering
{\includegraphics[width=7.0cm]{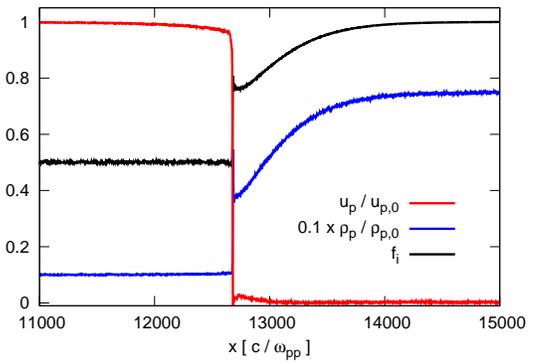}}
\par}
\caption{Shock structures averaged over the $y$ direction at $t=2000~{\Omega_{\rm cp}}^{-1}$. 
The red, blue, and black lines show the proton mean velocity normalized by the far upstream value, $u_{\rm p}/u_{\rm p,0}$, 
the proton density normalized by 10 times the far upstream value, $0.1\rho_{\rm p}/\rho_{\rm p,0}$, 
the ionization fraction, $f_{\rm i}$, respectively. }
\label{f3}
\end{figure}
\begin{figure}
{\centering
{\includegraphics[width=8.6cm]{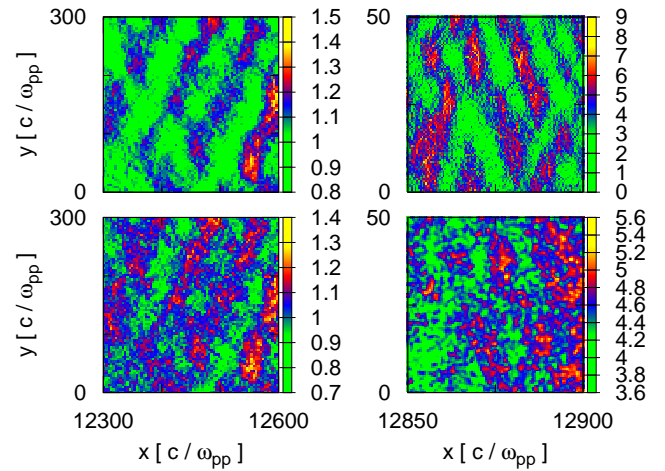}}
\par}
\caption{Magnetic field strength, $|B|/B_0$, (top) and density, $\rho_{\rm p}/\rho_{\rm p,0}$, (bottom) 
in the upstream (left) and downstream (right) regions at $t=2000~{\Omega_{\rm cp}}^{-1}$. 
Note that the spatial scale of the right figures is smaller than that of left.}
\label{f4}
\end{figure}
\begin{figure}
{\centering
{\includegraphics[width=8.6cm]{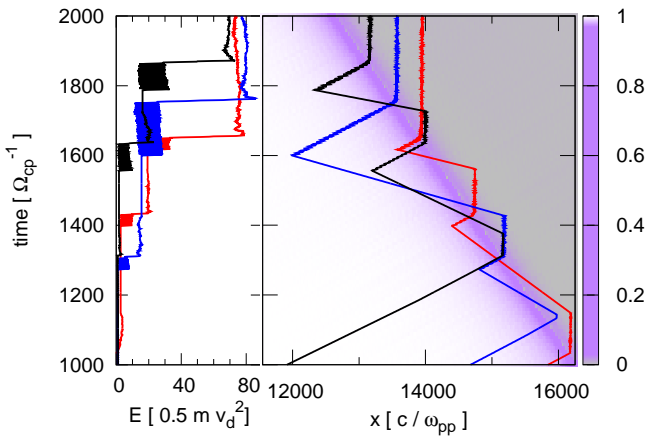}}
\par}
\caption{Trajectories of three accelerated particles (red, blue and black lines). 
The right panel shows the time evolution of three particle positions in $x$, 
where the background color shows the mean proton velocity, $u_{\rm p}/u_{\rm p,0}$. 
The left panel shows the time evolution of kinetic energies of the three particles.}
\label{f6}
\end{figure}

The phase space at time $t=2000~{\Omega_{\rm cp}}^{-1}$ is shown in Fig 1. 
The shock is located at $x=12700~c/\omega_{\rm pp}$ 
and propagating into the $-x$ direction 
with velocity $3.61~v_{\rm A}$ in the downstream rest frame, 
so that the shock velocity is $v_{\rm sh}=13.61~v_{\rm A}=2722~{\rm km/s}$ 
in the upstream rest frame and the total compression ratio is $r_{\rm tot}=3.77$. 
Note that if we redefine the Alfv{\'e}n velocity as 
$B_0/\sqrt{4\pi(\rho_{\rm p,0}+\rho_{\rm H,0})}$, 
the shock velocity becomes $19.25~v_{\rm A}$ in the upstream rest frame, 
where $\rho_{\rm H,0}$ is the upstream hydrogen mass density. 
The total compression ratio, $r_{\rm tot}=3.77$ is somewhat smaller than 
$3.93$ that based on the Rankine-Hugoinot relations for 
$v_{\rm sh}=19.25~v_{\rm A}$ and $\beta_{\rm p}+\beta_{\rm H}=1$. 
This is because the simulation box is two-dimensional space or 
because the behavior of pickup ions produced by ionization of 
hydrogen atoms is not that of gas with the adiabatic index of $5/3$.

Some hydrogen atoms leak into the upstream region from the downstream region. 
The leaking hydrogen atoms originate from hot hydrogen atoms produced by charge exchange 
between downstream hot protons and downstream hydrogen atoms. 
The number density of the leaking hydrogen atoms is 
about $7\%$ of that of upstream hydrogen atoms at the shock. 
The mean velocity of the leaking hydrogen atoms is 
$u_{x,{\rm leak}}=-0.23~v_{\rm sh}$ in the shock rest frame. 
The leaking hydrogen atoms are ionized by upstream electrons, 
protons and hydrogen atoms in the upstream region. 
Then, the ionized particles are picked up by the upstream flow and become pickup ions. 
In the upstream rest frame, the pickup ions become isotropic in $v_x-v_z$ plane by the magnetic field but 
the pickup ions and upstream protons do not relax to the same distribution. 
All the upstream protons are mainly thermalized at the collisionless shock at $x=12700~c/\omega_{\rm pp}$. 
In this simulation, the energy transfer from the pickup ion to the upstream plasma could be underestimated 
because all the reaction rates are artificially enhanced by a factor of $10^3$. 
On the other hand, upstream neutral particles freely penetrate the shock front without deceleration 
and are ionized in the downstream region. 
As the result, upstream neutral particles also become pickup ions in the downstream region.

In the left panel of Fig 2, we show the velocity distribution of hydrogen atoms 
in the downstream region, 
$12700~c/\omega_{\rm pp} \leq x \leq17000~c/\omega_{\rm pp}$, 
at time $t=2000~{\Omega_{\rm cp}}^{-1}$.
The velocity distributions have two components (narrow and broad) 
that are similar to expected from observed line profiles of H$\alpha$ emission \citep{chevalier78}.
The narrow component originates from upstream hydrogen atoms (before charge exchange), 
while the broad component originates from hot hydrogen atoms 
produced by charge exchange in the downstream region.

Fig 3 shows $y$-averaged shock structures at time $t=2000~{\Omega_{\rm cp}}^{-1}$.
The red, blue and black lines show the mean proton velocity of the $x$ direction, $u_{\rm p}/u_{\rm p,0}$, 
the proton density, $0.1\rho_{\rm p}/\rho_{\rm p,0}$, and ionization fraction, $f_{\rm i}$, respectively. 
It is well known that the shock thickness is about the gyro radius of protons 
for perpendicular shocks in fully ionized plasmas, 
that is, $\approx 10~c/\omega_{\rm pp}$ 
for $v_{\rm sh} \approx 10~v_{\rm A}$ \citep{leroy83}. 
For partially ionized plasmas, as shown in Fig 3, the velocity and density profiles 
have another scale length of the order of $10^3~c/\omega_{\rm pp}$ that 
corresponds to the ionization length scale. 
In the upstream and downstream regions, the plasma flow is gradually decelerated 
by the pressure of pickup ions produced in upstream and downstream regions. 
In this Letter, charge exchange and collisional ionization are enhanced 
by a factor of $10^3$ and $v_{\rm sh}\approx 10~v_{\rm A}$, 
so that the actual ionization length scale becomes about $10^7~c/\omega_{\rm pp}$ 
for young SNRs with $v_{\rm sh}\approx 10^2~v_{\rm A}$.
Furthermore, in the shock rest frame, the velocity jump at the subshock with the length scale 
of $10~c/\omega_{\rm pp}$ ($x=12700~c/\omega_{\rm pp}$) is $3.47$ 
and smaller than the total compression ratio, $r_{\rm tot}=3.77$. 
This is because the pickup ions produced in the upstream region make the Mach number small. 
The smaller velocity jump at the subshock makes the cosmic-ray spectrum soft and 
this can explain the observed gamma-ray spectra slightly steeper than the simplest prediction 
of the shock acceleration \citep{ohira12,blasi12}.

In Fig 4 we show the magnetic field strength and density structures in the upstream 
and downstream regions at time $t=2000~{\Omega_{\rm cp}}^{-1}$.
In the upstream region (left figures), the magnetic field strength (top) is correlated with the density (bottom). 
This fast magnetosonic mode might be excited by the Drury instability \citep{drury86} or other mechanisms. 
Detailed linear analyses will be addressed in future works.
Moreover, magnetic field structures of the $x$ and $z$ components, that are not shown in this Letter, 
show that the Alfv{\'e}n mode is also excited in the upstream region. 
There is the pressure anisotropy of pickup ions, $P_{\perp}/P_{\parallel}>1$, 
so that the Alfv{\'e}n mode is excited by the ion cyclotron instability \citep{raymond08}, 
where $P_{\perp}$ and $P_{\parallel}$ are pressures perpendicular and parallel to the magnetic field, respectively.
On the other hand, in the downstream region (right figures), the magnetic field strength (top) 
is anticorrelated with the density (bottom) and there is the pressure anisotropy of pickup ions, 
$P_{\perp}/P_{\parallel}>1$. 
Therefore, the downstream structure is due to the mirror instability \citep{raymond08}.
The pickup ions could excite other instabilities for parallel shocks \citep{ohira09}. 
Furthermore, denser regions and larger magnetic field fluctuations 
could be produced for higher Alfv{\'e}n Mach number shocks and the magnetic field could 
be amplified not only by plasma instabilities discussed above but also 
by turbulence \citep{giacalone07}.

The right panel of Fig 2 shows energy spectra of protons and hydrogen atoms in the downstream region, 
$12700~c/\omega_{\rm pp} \leq x \leq17000~c/\omega_{\rm pp}$, at time $t=2000~{\Omega_{\rm cp}}^{-1}$. 
Some protons are accelerated to about $10$ times the initial kinetic energy, $E_{\rm kin,0}=0.5mv_{\rm d}^2$. 
The mean relative velocity between the upstream flow and the leaking neutral particles 
is $v_{\rm rel,up}=1.67~v_{\rm d}=1.23~v_{\rm sh}$, so that  
when leaking hydrogen atoms are ionized and picked up by the upstream flow, 
their kinetic energy typically becomes $(v_{\rm rel,up}/v_{\rm d})^2~E_{\rm kin,0}$ in the upstream rest frame. 
When the pickup ions re-enter the downstream region, 
they are accelerated by adiabatic compression and their energy becomes 
$r_{\rm tot}(v_{\rm rel,up}/v_{\rm d})^2~E_{\rm kin,0}\approx 10~E_{\rm kin,0}$.
Because the cross section of charge exchange steeply decreases 
with the relative velocity 
for $v_{\rm rel}>3000~{\rm km/s}$, the second neutralization of 
accelerated particles is rare in this Letter. 
For slower shock velocity ($v_{\rm sh} < 2000~{\rm km/s}$), 
the multiple neutralization of accelerated particles 
and more leakage of hydrogen atoms can be expected.
The total kinetic energy of accelerated particles is nearly $10~\%$ 
of the total kinetic energy of all particles, 
so that the temperature of thermal component becomes somewhat 
lower than that for fully ionized plasmas.

Trajectories of representative accelerated particles are shown 
in the right panel of Fig 5. 
The mean proton velocity, $u_{\rm p}/u_{\rm p,0}$, is shown 
by the background color, where the white and gray regions show 
upstream and downstream regions and the purple region shows the precursor region. 
The left panel of Fig 5 shows the time evolution of kinetic energies of 
the representative accelerated particles. 
For example, in the case of the back line, the particle interacts with the shock at $t\approx1300~{\Omega_{\rm cp}^{-1}}$, and becomes a hydrogen atom by charge exchange at $t\approx1400~{\Omega_{\rm cp}^{-1}}$ and returns back to the upstream region.  At $t\approx1550~{\Omega_{\rm cp}^{-1}}$, it is ionized and picked up by the upstream flow, and accelerated. After that, the pickup ion experiences the shock heating again at $t\approx1600~{\Omega_{\rm cp}^{-1}}$. 
After that, the particle repeats these processes again. 
These processes can be regarded as injection to the shock acceleration. Therefore neutral particles could be important for injection into the shock acceleration \citep{ohira10,ohira12}. 

In conclusion, we have investigated nonrelativistic collisionless perpendicular shocks 
propagating into partially ionized plasmas by a new hybrid simulation that 
solves ionization of hydrogen atoms, particle motions and Maxwell's equations. 
We have found the followings: 
1) Nearly $10\%$ of hydrogen atoms leak 
into the upstream region from the shock downstream region.
2) The leaking hydrogen atoms become pickup ions in the upstream region 
and they are preferentially accelerated by the shock. 
3) The accelerated pickup ions decrease the temperature. 
4) The pickup ions modify the shock structure and excite plasma instabilities 
in the upstream and downstream regions. 
Hence, the ionization fraction could be relevant to 
the injection efficiency of the shock acceleration, 
the spectral index of accelerated particles, magnetic field strength 
and the temperature of the thermal component. 
In addition, we have found that the velocity distributions of 
hydrogen atoms in the downstream region have narrow and broad components.

Above quantitative values should depend on the shock Mach number, 
the ionization fraction, the shock velocity, the density, the magnetic field orientation, 
and so on. 
We have specified neutral particles as hydrogen atoms in this Letter. 
Because helium has about $25\%$ of the shock kinetic energy, 
effects of helium atoms are also important. 
The ionization fraction of helium in the upstream region depends on time 
because helium atoms are ionized by radiation from the downstream region \citep{ghavamian00}. 
Therefore, the injection of helium ions into the shock acceleration could depend on the age 
of SNRs. 
The cosmic-ray injection history of helium ions is important to understand the spectrum of 
cosmic-ray helium \citep{ohira11}. 
We have not solved electron dynamics in this Letter. 
As with the pickup ions, knock-on electrons produced 
by collisional ionization of leaking neutral particles have a large velocity, 
so that the knock-on electrons are a promising candidate for injection particles 
into the shock acceleration. 
These issues will be addressed in future work. 

Numerical computations were carried out on the XC30
system at the Center for Computational Astrophysics (CfCA)
of the National Astronomical Observatory of Japan.
The author thanks T. Inoue and R. Yamazaki for useful comments. 
This work is supported by Grants-in-aid from the Ministry of Education, 
Culture, Sports, Science, and Technology (MEXT) of Japan, No.~24$\cdot$8344.

\end{document}